\documentclass[11pt]{article}
\usepackage{amsmath,amsfonts, amssymb,graphicx,geometry,authblk,rotating,xcolor, mathtools} 
\usepackage[ruled]{algorithm2e}
\usepackage{multirow}
\usepackage{url}

\title{A learning-based multiscale method and its application to \\ inelastic impact problems} 
\author{Burigede Liu$^*$, Nikola Kovachki$^*$, Zongyi Li$^*$, Kamyar Azizzadenesheli$^\dagger$, \\Anima Anandkumar$^*$, Andrew Stuart$^*$, Kaushik Bhattacharya$^*$}
\affil{$^*$Department of Engineering and Applied Science, California Institute of Technology, Pasadena CA 91011}
\affil{$^\dagger$Department of Computer Science, Purdue University, West Lafayette IN 47906}
\date{}

\begin{document}
\maketitle
{\bf
The macroscopic properties of materials that we observe and exploit in engineering application result from complex interactions between physics at multiple length and time scales: electronic, atomistic, defects, domains  etc.  Multiscale modeling seeks to understand these interactions by exploiting the inherent hierarchy where the behavior at a coarser scale regulates and averages the behavior at a finer scale.  This requires the repeated solution of computationally expensive finer-scale models, and often \emph{a priori} knowledge of those aspects of the finer-scale behavior that affect the coarser scale (order parameters, state variables, descriptors, etc.).  We address this challenge in a two-scale setting where we learn the fine-scale behavior from off-line calculations and then use the learnt behavior directly in coarse scale calculations.  The approach draws from recent successes of deep neural networks, in combination with ideas from model reduction. The approach builds on the recent success of deep neural networks by combining their approximation power in high dimensions with ideas from model reduction. It results in a neural network approximation that has high fidelity, is computationally inexpensive,  is independent of the need for \emph{a priori} knowledge, and can be used directly in the coarse scale calculations.  We demonstrate the approach on problems involving the impact of magnesium, a promising light-weight structural and protective material.
}

\paragraph{Significance}
The development and optimization of new materials is challenging because the macroscopic behavior of materials is the result of mechanisms operating over a wide range of length and time scales. Traditional empirical models are computationally inexpensive, but are unable to describe this complexity of behavior. At the other end, high-fidelity concurrent multiscale methods replace the need for empirical information with first principles modeling but are often prohibitively expensive. We propose a method that provides the fidelity of concurrent multiscale modeling and beyond, at a few times the computational cost of an empirical model, by using machine learning combined with model reduction to approximate the solution operator of the fine scale model.



\newpage
The macroscopic behavior of materials is the end result of mechanisms operating over a wide range of length and time scales, where mechanisms at the larger scales both filter (average) and modulate (set the boundary condition) those at the lower scales \cite{p_book_01}.  The development and optimization of new material/structural systems therefore require an understanding of the various mechanisms and their interactions across the scales.  
While each mechanism has been studied by developing models at an individual scale: density functional theory at the electronic scale  \cite{g_book_14}, molecular dynamics at the atomistic scale \cite{f_book_10}, defect models at the nanoscale \cite{bc_book_13}, crystal plasticity at the sub-grain scale \cite{asaro1983crystal}, empirical inelastic theories at the engineering scale \cite{gfa_book_13} etc., recent work has focused on multiscale modeling that seeks to understand the behavior across multiple scales \cite{f_book_09,dr_book_11}.   The entire range of material behavior is first divided into a hierarchy of scales \cite{van2020roadmap}, the relevant mechanisms at each scale are identified and analyzed using theories/tools based on an individual scale, and the hierarchy is stitched together by passing information between scales.  
While the mathematical theory of homogenization \cite{bensoussan2011asymptotic,pavliotis2008multiscale} provides a concrete basis in specialized situations, the underlying conceptual framework has been adopted broadly.  Importantly, multiscale modeling has explained experimental observations where empirical models have failed (e.g. strength of solids in extreme conditions \cite{betal_jap11}).

One widely used approach is the sequential multiscale or parameter passing method which extends the empirical approach by evaluating parameters in the coarse model using information from lower scale models\cite{weinan2011principles}.  Examples include training atomistic models from first principles \cite{cheng2019first}, inferring defect kinetics from atomistic simulations 
\cite{fu2005multiscale}, and fitting macroscopic plasticity models from crystal plasticity calculations \cite{balasubramanian2002elasto}.  While the coarse model can be derived in some situations (e.g. linear elasticity \cite{bensoussan2011asymptotic}), it has to be postulated \emph{a priori} in most situations.   Another approach, with greater fidelity, is the concurrent multiscale method that evaluates mechanisms operating at different scales in parallel so that the small and large scale models are computed concurrently.  Examples include the Car-Parrinello molecular dynamics \cite{car1985unified}, the quasicontinuum method  \cite{tadmor1996quasicontinuum}, and the FE$^2$ approach 
\cite{feyel2000fe2}.  However, concurrently evaluating the mechanisms across scales is expensive and can exceed the present computational power for analyzing practical engineering problems.  Further, it is often necessary to postulate  \emph{a priori} the descriptors (state or internal variables, order parameters) by which the coarser and finer scale models communicate. The existence and identification of such descriptors are far from clear, especially in time-dependent phenomena \cite{b_prs_99}.

In short, the practical implementation of the multiscale modeling of materials suffers from two challenges.  The first is that one often needs  \emph{a priori} or empirical knowledge about the interaction between models at various scales.  The second, especially in the concurrent multiscale approach, is the need to repeatedly solve the expensive finer scale model only to use a very small portion of the information.  This naturally raises the question:
\emph{how can the data generated by repeatedly solving the finer scale model be utilized to create a computationally efficient surrogate of its solution operator, that can directly be used at the coarser scale with no further modeling?}

Machine-learning and especially deep neural networks have been extremely successful in image recognition \cite{lecun1995convolutional, he2016deep} and natural language processing tasks \cite{goldberg2017neural, collobert2008unified}.   There is also a growing literature on the use of these methods in materials science \cite{kd_armr_15}.  Machine learning has been combined with theoretical calculations, combinatorial synthesis, and high through-put characterization to rapidly identify materials with desired properties \cite{l_ncm_19,umehara2019analyzing,jetal_aplm_13}.
It has also been applied to parameter passing \cite{marchand2020machine, cole2020machine,wen2019hybrid} and 
to the inversion of experimental data \cite{m_mse_20,cd_mse_18}.
Somewhat closer to our work, image classification and natural language processing have been applied to approximate material constitutive behavior \cite{Mozaffar2019DeepPlasticity, jordan2020neural} 
and  homogenization of material behavior \cite{liu2019deep1, xiao2019machine}.

In this work, we develop a framework to answer the question raised above in a two-scale setting.  Specifically we demonstrate, on problems involving the impact of a polycrystalline inelastic solid, that it is possible to solve macroscopic problems with the fidelity of concurrent multiscale modeling and beyond (because we do not need \emph{a priori} identification of state variables) at few times the computational cost of solving the problem with an empirical model.   A critical challenge is that material models are described as partial differential equations that map inputs from one function space (e.g. average strain history) to outputs on another function space (correspondingly, the resulting stress history).   While typical approaches use a finite-dimensional subspace obtained by discretization to solve these problems, it is desirable for the learnt map to be independent of the particular discretization or resolution.  Therefore we use the approach developed in Bhattacharya {\it et al.} \cite{bhattacharya2020model} that combines model reduction and neural networks for high-fidelity approximations of maps between function spaces.

\section{Broad overview of our approach}  \label{sec:over}

Consider a heterogeneous body occupying the region $\Omega \subset {\mathbb R}^d, d=2,3$ in the reference configuration.  We are interested in situations where the ratio $\varepsilon$ of the scale of the heterogeneity to that of the body is small.  Let $u:\Omega \to {\mathbb R}^d$ denote the deformation and $F = \nabla u$ the deformation gradient.  The state of the body is described by a set of internal variables $\xi: \Omega \to {\mathbb R}^m$ and the deformation gradient.  The constitutive relation is described by the (Piola-Kirchhoff) stress function $S^\varepsilon: {\mathbb R}^{d\times d} \times {\mathbb R}^{m} \times\Omega \to {\mathbb R}^{d\times d}$ and a kinetic relation $K^\varepsilon: {\mathbb R}^{d\times d} \times 
{\mathbb R}^{m} \times {\mathbb R}^{m} \times\Omega \to {\mathbb R}^{m}$ that describes the evolution of the internal variables.   Let $\rho^\varepsilon: \Omega \to {\mathbb R}$ denote the (referential) density.

Given $u_0, v_0, u^*, s^*$, the displacement $u^\varepsilon$ and internal variables $\xi^\varepsilon$ are given by the solution of the system
\begin{align}
& \nabla \cdot S^\varepsilon  = \rho^\varepsilon u^\varepsilon_{tt} && \text{on } \Omega \label{eq:mb}\\
&K^\varepsilon = 0 && \text{on } \Omega \label{eq:kr}\\
&u^\varepsilon(x,0) = u_0(x), \quad u^\varepsilon_t(x,0) = v_0(x), \quad \xi^\varepsilon(x,0)= \xi_0&& \text{on } \Omega \label{eq:ic}\\
& u^\varepsilon(x,t) = u^*(x,t) &&  \text{on } \partial_1\Omega \label{eq:dbc}\\
&S^\varepsilon (\nabla u^\varepsilon, \xi^\varepsilon,x)n(x) = s^*(x,t) &&  \text{on } \partial_2\Omega \label{eq:tbc}
\end{align}
where $\partial_1\Omega \cup \partial_1\Omega = \partial \Omega$.  
(\ref{eq:mb}) is the equation of motion, (\ref{eq:kr}) the kinetic relation that describes the evolution of the internal variables, (\ref{eq:ic}) the initial condition and (\ref{eq:dbc}), (\ref{eq:tbc}) the boundary conditions.  Note that the displacement and internal variables oscillate on a scale smaller than $\varepsilon$ and we emphasize this with the superscript.  In this work, we consider an almost periodic medium where $S^\varepsilon(F,\xi,x) = S(F,\xi,x,x/\varepsilon)$, $K^\varepsilon(F,\xi, \xi_t, x) = K(F,\xi, \xi_t, x,x/\varepsilon)$, $\rho^\varepsilon (x) = \rho(x,x/\varepsilon)$ where $S:  {\mathbb R}^{d\times d} \times {\mathbb R}^{m} \times\Omega\times Y  \to {\mathbb R}^{d\times d}$, $K:{\mathbb R}^{d\times d} \times {\mathbb R}^{m} \times {\mathbb R}^{m} \times\Omega \times Y \to {\mathbb R}^{m}$, 
$\rho: \Omega \times Y \to {\mathbb R}$ are periodic with period $Y$ ($|Y|$ = 1) in their last variable.

We now show that the  solution to this problem which has to be resolved on the fine scale $\varepsilon$ may be approximated with that of the \emph{macroscopic problem}
\begin{align}
& \nabla \cdot \bar S  = \bar{\rho} u_{tt} && \text{on } \Omega \label{eq:mmb}\\
&u(x,0) = u_0(x), \quad u_t(x,0) = v_0(x) && \text{on } \Omega \label{eq:mic}\\
& u(x,t) = u^*(x,t) &&  \text{on } \Gamma_1 \label{eq:mdbc}\\
&\bar{S} n(x) = s^*(x,t) &&  \text{on } \Gamma_2 \label{eq:mtbc}
\end{align}
 where the stress $\bar{S}$ and displacement $u$ are smooth on the scale of $\varepsilon$, for an appropriate \emph{macroscopic constitutive behavior or closure relation} 
\begin{equation} \label{eq:mapfs}
{\mathcal F}: \{F(\tau): \tau\in (0,t) \} \mapsto \bar{S} (t) \quad t \in (0,T)
\end{equation}
that describes how the macroscopic stress $\bar S$ depends on the history of the macroscopic deformation gradient consistent with the fine scale problem.

To do so, we rewrite (\ref{eq:mb},\ref{eq:dbc},\ref{eq:tbc}) as
\begin{equation}
\int_\Omega S^\varepsilon \cdot \nabla w dx - \int_{\partial_2 \Omega} s^* \cdot w = \int_\Omega \rho^\varepsilon u^\varepsilon_{tt} \cdot w dx  \quad \forall \ w
\end{equation}
and make the two-scale ansatz\footnote{We ignore the displacement boundary condition (\ref{eq:dbc}) to simplify the treatment, but can incorporate it  using a boundary layer.}
\begin{align}
u^\varepsilon(x,t) & = u^0(x,t) + \varepsilon u^1(x,x/\varepsilon,t) + \varepsilon^2 u^2 (x,x/\varepsilon,t) + \dots\\
\xi^\varepsilon(x,t) & = \xi^0(x,x/\varepsilon,t) + \varepsilon \xi^1(x,x/\varepsilon,t) + \dots
\end{align}
where $u^j(x,y,t)$ is periodic in \(y\) with period $Y$ for any \(j=1,2,\dots\).   Note that
\begin{equation}
S^\varepsilon = S(\nabla_x u^0+\nabla_y u^1, \xi^0,x,y) + \dots, \quad
K^\varepsilon =  K(\nabla_x u^0+\nabla_y u^1, \xi^0, \xi^0_t, x,y) + \dots.
\end{equation}
Taking a test function of the form
\begin{equation}
w(x,t) = w^0(x,t) + \varepsilon w^1(x,x/\varepsilon,t) + \varepsilon^2 w^2 (x,x/\varepsilon,t) + \dots
\end{equation}
where $w^j(x,y,t)$ is periodic in \(y\) with period \(Y\) for any \(j=1,2,\dots\), we obtain
\begin{equation}
\int_\Omega S^\varepsilon \cdot (\nabla_x w^0 + \nabla_y w^1) dx - \int_{\partial_2 \Omega} s^* \cdot w^0 = \int_\Omega \rho^\varepsilon u^0_{tt} \cdot w^0 dx \quad \forall \ w^0,w^1.
\end{equation}
Integrating this over $Y$, we obtain
\begin{align}
&\int_\Omega \langle S^\varepsilon \rangle \cdot \nabla_x w^0 dx - \int_{\partial_2 \Omega} s^* \cdot w^0 = \int_\Omega \langle \rho^\varepsilon \rangle w^0_{tt} \cdot w^0 dx  &&\forall \ w^0, \label{eq:macro}\\
&\int_\Omega \int_Y S^\varepsilon \cdot \nabla_y w^1 dy \ dx = 0  &&\forall \ w^1 \label{eq:second}
\end{align}
where $\langle \cdot \rangle$ denotes the average over the unit cell $Y$.  The first equation  (\ref{eq:macro}),  initial conditions (\ref{eq:ic})$_{2,3}$, and boundary conditions (\ref{eq:dbc},\ref{eq:tbc}) define the macroscopic problem (\ref{eq:mmb})-(\ref{eq:mtbc}).  We treat $x$, $F = \nabla_x u^0$, and $t$ as parameters in the second equation (\ref{eq:second}) and obtain the \emph{unit cell problem}:
\begin{align}
& \nabla \cdot S(F+\nabla v, \xi,x,y)  = 0 && \text{on } Y \label{eq:ueq}\\
&K (F+\nabla v, \xi,\xi_t,x,y) = 0 && \text{on } Y\label{eq:ukr}\\
& \xi(y,0) = \xi_0(y) && \text{on } Y \label{eq:uic}\\
&v  \ \text{periodic} \label{eq:ubc}.
\end{align}
Now, given $\xi_0$,  the solution to this unit cell problem defines the requisite macroscopic closure relation (\ref{eq:mapfs}) with $\bar{S} = \langle S^\varepsilon \rangle$ for the macroscopic problem.

It is common to specify the stress function in terms of the Cauchy stress $\sigma = (\text{det } F)^{-1} S F^T$ which is symmetric. Further, the underlying physical models $S,K$ are invariant under a change of frame and so is the map (\ref{eq:mapfs}).  It follows that ${\mathcal F}: \{R(\tau) F(\tau), \tau \in (0,t)\} \mapsto R(t) \langle S \rangle (t)$ for any time-dependent rotation $R(t)$.  According to the polar decomposition theorem, the deformation gradient $F = R U$ where $R$ is a rotation and $U$ is positive definite-symmetric.  Therefore, it suffices to specify the equivalent \emph{constitutive behavior or closure relation}
\begin{equation} \label{eq:fbar}
\bar{\mathcal F}: \{ U(\tau): \tau \in (0,t) \} \mapsto \langle \sigma \rangle (t) \quad t \in (0,T).
\end{equation}
\vspace{0.1in}

Now, the implementation of the multiscale problem above requires the calculation of the map $\bar{\mathcal F}$, and therefore the unit cell problem at each macroscopic point $x$ and at each instant $t$; this is extremely expensive.  Our idea is to \emph{ learn the macroscopic constitutive behavior}  using model reduction and deep neural networks following the approach of Bhattacharya {\it et al.} \cite{bhattacharya2020model} by utilizing \emph{ data generated by solutions of the unit cell problem} over various strain histories obtained from an appropriate probability distribution in the space of strain histories.  To do so, we observe that the unit cell problem in fact specifies the map 
\begin{equation} \label{eq:map}
\Psi: \{ U(t): \tau \in (0,T) \} \mapsto \{ \langle \sigma \rangle (t): t \in (0,T)\}.
\end{equation}
for a some $T>0$.  Further, the underlying equations and therefore this map is \emph{causal}, i.e., $\langle S \rangle (t)$ depends only on $\{F(\tau): \tau\in (0,t) \}$.   Finally, any map of the form (\ref{eq:map}) that is causal uniquely defines a map of the form (\ref{eq:fbar}).  Therefore, we learn $\Psi$ which maps one function to another.


The map $\Psi: L_2 ( (0,T); {\mathbb R}^{d(d+1)/2}) \to  L_2 ( (0,T); {\mathbb R}^{d(d+1)/2})$ is one between infinite dimensional Hilbert spaces.  However, the data is discretized and standard neural networks are defined as maps between finite-dimensional spaces.  Thus we seek a finite dimensional approximations of these infinite-dimensional spaces. Further, we want our approximation and our architecture to be independent of any specific discretization.  To that end, we seek maps $p_i: L_2 \to {\mathbb R}^{d_i}$ and $p_o: L_2 \to {\mathbb R}^{d_o}$ that reduces (project) the input and output spaces and maps $\ell_i:  {\mathbb R}^{d_i} \to L_2$ and $\ell_o: {\mathbb R}^{d_o} \to L_2$ that lift them back up
such that $p_i \circ \ell_i \approx id$, $p_o \circ \ell_o \approx id$.  We then find an approximate map $\psi:  {\mathbb R}^{d_i} \to  {\mathbb R}^{d_o}$ such that 
\begin{equation} \label{eq:la}
\Psi \approx \ell_o \circ \psi \circ p_i.
\end{equation}

In this work, we use principal component analysis (PCA) \cite{wold1987principal} to specify the maps $p_i, p_o$ and a fully-connected deep neural network with $M$ layers to approximate $\psi$: 
\begin{equation}
\label{eq:encoder_decoder_output} 
\psi(s) = W_M...\omega(W_2\omega(W_1(s)+b_1)+b_2)...+b_M, s\in \mathbb{R}^{d_i}, 
\end{equation} 
where $W_1,...,W_M$ are the weight matrices, $b_1,...,b_M$ are the bias vectors, and $\omega$
is the scaled exponential linear unit (SELU) \cite{Klambauer2017Self-normalizingNetworks} with scaling constants $\chi = 1.67$ and $\beta = 1.05$. We approximate the PCA specified maps by a standard SVD algorithm and the weight and bias parameters of the neural network with standard stochastic gradient based minimization techniques \cite{bhattacharya2020model}.

This learnt approximate map replaces the constitutive relation in a macroscopic integrator.  The approach is summarized in Algorithm \ref{al:overview}.

\begin{algorithm}[t]
\caption{Overview of the learning based multiscale modeling.  \label{al:overview}}
{\bf Off-line (once for any material)}  \\
Solve the unit cell problem for various trajectories $U(t), t\in(0,T)$\\
Use this data to train the approximation (\ref{eq:la})\\
\vspace{0.2in}
{\bf Online (for each simulation)}\\
{\it Initialize}\\
Discretize the macroscopic problem in space using finite elements.\\
Discretize the macroscopic problem in time and use an explicit macroscopic integrator.\\
\While{$t_n \le T$}{
\ \ \ \ At each quadrature point:\\
\ \ \ \ \ \ \ \ The integrator provides the deformation history $\{F_m\}_{m\le n}$.\\
\ \ \ \ \ \ \ \ Use the polar decomposition theorem $F_m = Q_m U_m, m = 1, \dots, n$.\\
\ \ \ \ \ \ \ \  Generate a trajectory $U(t)$ consistent with the history $\{u_m\}_{m\le n}$.\\
\ \ \ \ \ \ \ \  Use the learnt approximation (\ref{eq:la}) to obtain the Cauchy stress history $\langle \sigma \rangle (t)$.\\
\ \ \ \ \ \ \ \  Return the approapriately oriented  stress $Q_n \langle \sigma \rangle (t_n) Q_n^T$ to the integrator.\\
\ \ \ \  Update the deformation using the macroscopic integrator.
}
\end{algorithm}

\paragraph{Remarks}
\begin{enumerate}

\item  The proposed approach does not require any explicit macroscopic constitutive relation.  Instead, the constitutive behavior is implicitly defined by the unit cell problem  (\ref{eq:ueq}) - (\ref{eq:ubc}).  We learn the solution of this problem using a neural network, and this neural net acts as the constitutive relation in macroscopic problems.  Similarly the proposed approach does not require any macroscopic internal variable (descriptor of the macroscopic state of the material) \cite{rice1971inelastic}.  These aspects are distinct from almost all other approaches, traditional and multiscale.

In the \emph{traditional approach}, we do not consider the full problem but postulate the existence of an empirical \emph{macroscopic internal variable} $\zeta : \Omega \to {\mathbb R}^{m'}$ and macroscopic constitutive relations $\tilde {S}: {\mathbb R}^{d\times d} \times {\mathbb R}^{m'} \times\Omega \to {\mathbb R}^{d\times d}$, 
$\tilde{K}: {\mathbb R}^{d\times d} \times {\mathbb R}^{m'} \times {\mathbb R}^{m'} \times\Omega \to {\mathbb R}^{m'})$, and to solve the macroscopic problem (\ref{eq:mmb})-(\ref{eq:mtbc}) supplemented with the macroscopic kinetic relations
$\bar{K} = 0,  \  \zeta(x,0) = 0$ on $\Omega$.
It is computationally inexpensive, but relies on limited empirical knowledge and makes no explicit use of the micro-scale physics.

Such an approach can be justified using homogenization theory in linear elasticity and with limitations (away from long wavelength instabilities) in finite elasticity, i.e. in theories where we do not have any internal variables.  However, the rigorous justification remains an open problem in models with internal variables including plasticity.  Indeed, the strict statement, in particular, the existence of a properly defined macroscopic internal variable, is likely false with pinning, memory effects, mixing of energetic and kinetic terms etc., but may hold in some approximate sense.  Further, even if the original full problem depends on the rate of change of the internal variable, the macroscopic constitutive functions may depend on the history of the deformation.

In the \emph{parameter passing multiscale approach}, we solve the unit cell problem under various conditions and use the solutions to complement empirical data in the development of $\tilde{S}, \tilde{K}$.  This is an exercise in regression, and well-suited for machine learning \cite{liu2019deep1}.  
This approach can be relatively computationally inexpensive and it incorporates some microscale physics.  However, it still relies on limited empirical knowledge and postulates the existence and identification of the macroscopic internal variable (descriptors) $\zeta$ and constitutive functions $\tilde{S}, \tilde{K}$.   

In the \emph{concurrent multiscale approach}, we integrate the macroscopic problem as in our proposed method, but  solve the unit problem at each quadrature point for every point in the time interval of interest with initial data $\xi_0$ based on the macroscopic internal variable $\zeta$.  Thus, it does not postulate the existence of $\tilde{S}, \tilde{K}$, but requires the \emph{a priori} identification of a macroscopic internal variable $\zeta$ and the relation between $\xi$ and $\zeta$.   This can be challenging because the approach entails the `inverse homogenization' problem of generating a representative state of the microscopic internal variables given a macroscopic internal variable.  Although the method makes extensive use of the miccroscopic physics with little need for empirical input, it is extremely computationally expensive. 

\item We can, in principle, learn $\Psi$ directly from experimental data rather than the solution of a lower scale model on a unit cell if sufficient experimental data were available.
Similarly, we do not need to solve the unit cell problem, but use an approximate method at that scale.  We demonstrate this later by replacing the unit cell problem with a Taylor model i.e. assuming that $v=0$ in the unit cell problem.

\item We use PCA for our model reduction $p_i, p_o$ and a deep neural network for the approximate map $\Psi$.  We can replace these with other model reduction (e.g. auto-encoders) and machine learning architectures (e.g. convolutional networks, random features).  

\item The proposed approach requires that we train a neural network over the entire duration $T$ of any simulation, and this may be unknown {\it a priori}.  There are two reasons for this requirement.  First, we do not carry any information about the state of the material, but only the average deformation gradient versus the average stress.   Second, we do not make any assumption on fading memory.  Learning these aspects from the data gathered by the simulations of the unit cell is interesting, but beyond the scope of the current work.

\item 
The cost of the approach has two components. The first is the one-time off-line cost of generating the data and training the neural network.  This typically scales linearly with the period over which the neural network needs to be trained, the number of simulations in the training data set, and the number of epochs of training required.  The latter two may also depend on the first, and this remains to be understood. In the examples we study, the number of simulations required to train a neural network is similar to the number of quadrature points in a typical sample.  For this reason, our off-line cost is similar to that of a single simulation of the concurrent multiscale approach.
The second is the online cost of evaluating the neural network during the simulation. This is typically small compared to the cost of the macroscopic time integration, but can scale quadratically with the period of the simulation since the evaluation of the stress at any instant requires the evaluation of the entire trajectory. Our examples show that the cost of the evaluation is a few times the cost of evaluating an empirical constitutive relation; however, we are able to take large time steps with the macroscopic integrator since we know the entire trajectory in our approach.

\item 
Our approach requires us to store the history of deformation. However, since we use model reduction, we do not need to store the trajectory for every time step but only require sampling it sufficiently to reconstruct the history.  This enables us to manage memory.
\end{enumerate}

\section{Learning crystal plasticity} \label{sec:lcp}

We demonstrate our approach by studying the inelastic deformation of polycrystalline solids.  

\subsection{The unit cell problem: crystal plasticity with twinning} \label{sec:cp}
A polycrystal is a medium made of a collection of disjoint subdomains or grains.  Each grain is made of the same material, but the orientation of the grain may differ with respect to a reference frame.   
We specify the texture (number, size and orientations) of the grains in the unit cell by a (piece-wise) rotation-valued orientation function $Q: Y \to SO(d)$.  The behavior of the material at a point $y$ is given by that of a reference material rotated by $Q(y)$: $S(F,\xi,x,y) = S_r(FQ(y), \xi_{Q(y)}), 
K(F,\xi,\dot{\xi},x,y) = K_r(FQ(y), \xi_{Q(y)}, \dot{\xi}_{Q(y)})$  where $\xi_{Q}$ is the action of the rotation $Q$ on the internal variable $\xi$, and $S_r, K_r$ describe the behavior in the reference material.

Crystalline solids can undergo plastic or inelastic deformation governed by slip on one of $n_s$ slip systems and twinning on one of $n_t$ twin systems. The kinetic relation $K_r$ describes how the internal variables -- the crystallographic orientation $Q$, the total inelastic deformation gradient $F_\text{in} \in {\mathbb R}^{d\times d}$, the slip activity $\gamma = \{\gamma_\alpha\}_{\alpha=1}^{n_s}$ in the $n_s$ slip systems, and the twin volume fractions $\lambda = \{\lambda_\beta \}_{\beta=1}^{n_t}$ in the $n_t$ twin systems that satisfy $\lambda_\beta \in [0,1], \ \sum_\beta \lambda_\beta = 1$ -- evolve.  The details of the model following Chang and Kochmann \cite{Chang2015AMagnesium} are provided in \ref{app:cp}.
We study two versions of this model.

\paragraph{2DFFT}
The first is in two dimensions, has two slip systems and no twinning i.e., $n_s=2, n_t = 0$ (details in \ref{app:cp}).  The initial texture consists of 16 grains generated using periodic Voronoi tessellation \cite{fritzen2009periodic}.  The corresponding full-field unit cell problem is solved using a fast Fourier transformation scheme following Vidyasagar et al.  \cite{vidyasagar2017predicting}.

\paragraph{3DTaylor}
The second is in three dimensions and motivated by magnesium which is of current interest as a lightweight structural material. The detailed slip and twin systems, and the associated parameters are given in \ref{app:cp}.   We do not solve the unit cell problem, but use the Taylor averaging assumption that the deformation gradient is uniform in the unit cell ($v=0$)  \cite{ketal_book_00}.   We use an initial texture of 128 randomly oriented grains. 

\subsection{Learning crystal plasticity}

The first task is to generate the data, and this requires sampling $L_2((0,T);{\mathbb R}^{d(d+1)/2})$.  We seek a strain path that is smooth but changes direction arbitrarily. To that end, we divide $(0,T)$ into $N$ intervals $\Delta t_n = t_n - t_{n-1}, n = 1, \dots N$ where $0 = t_0 < t_1 < \dots < t_N=T$ and set 
$(U_{ij})_n = (U_{ij})_{n-1} + \nu_nU_{max}\sqrt{\Delta t_n},$ $i,j = 1, \dots, d, i \le j
$
where $\nu_n \in \{-1,1\}$ follow a Rademacher distribution.  
We take $U_{ij}(t)$ to be the cubic  Hermite interpolation of $\{ (t_n, (U_{ij})_n) \}$.  
We take $N=10$ and study both fixed and random time intervals.

\begin{figure}[t]
    \centering
    \includegraphics[width=5in]{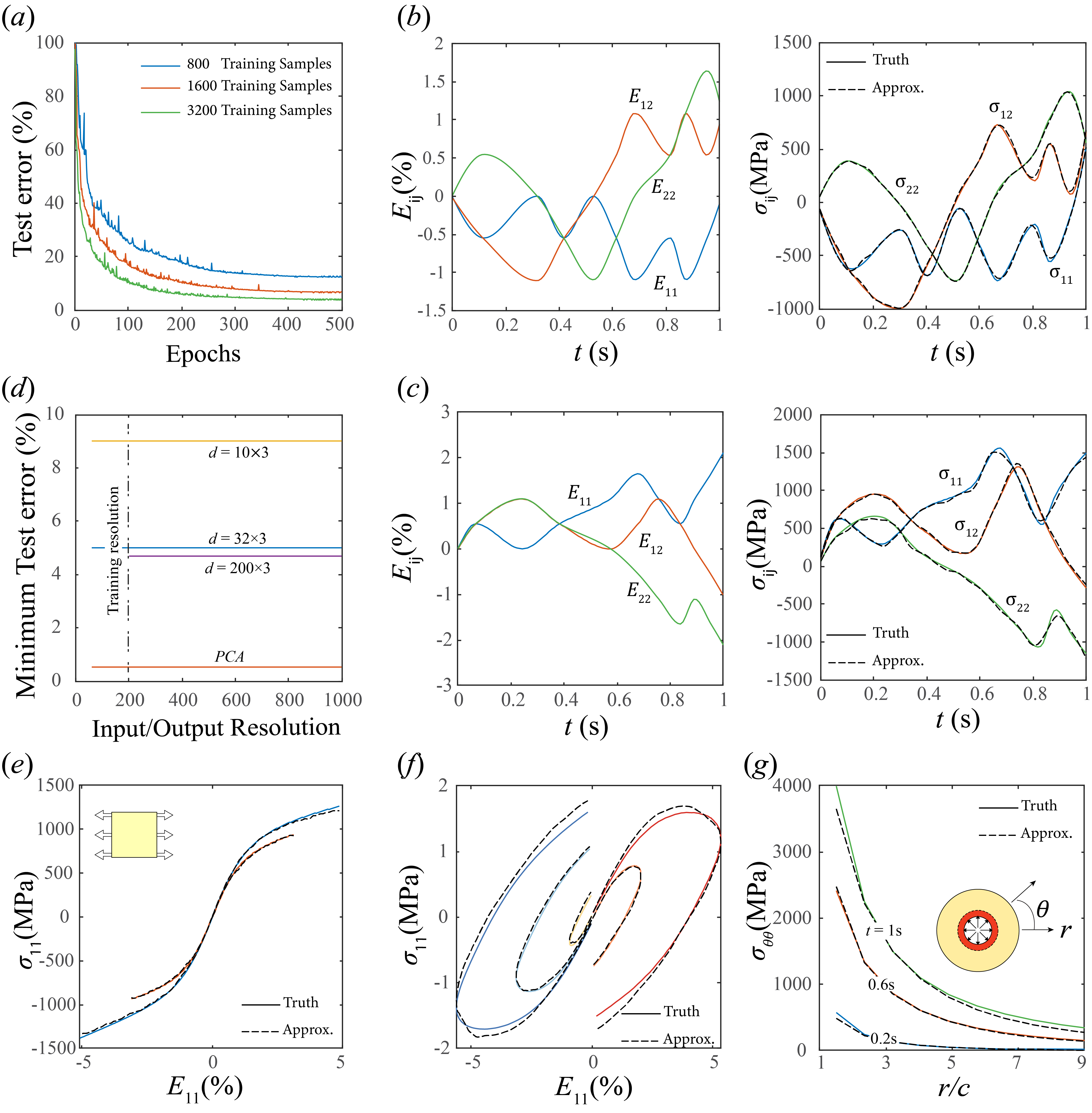}
    \caption{Deep learning approximation of 2D crystal plasticity.  (a) Average test error  for various training sample sizes and training epochs. (b) Typical sample from training set. (c) Typical sample from test set, (d) Test error vs. resolution of the input and output.  (e) Test for uniaxial strain loading.  (f) Test for uniaxial strain load-unload.  (g) Test for shear (cavity expansion).   
    }
    \label{fig:2dfftresult}
\end{figure}

We begin with  2DFFT.    We generate 6000 strain paths $U(t)$ using random time steps and solve the unit cell problem for the average stress $\langle\sigma\rangle(t)$ for each of these paths using a spatial resolution of $64 \times 64$ and 1000 time steps.  We then down sample the data to 200 time steps.   A single sample of data consists of the pair $\{U(t),\langle \sigma \rangle(t)\}$.  We reserve 2000 samples for testing and use various parts of the remaining 4000 for training.  We use a PCA dimension of $32\times 3$ for both the input and output spaces (here  $d(d+1)/2=3$).  We define the test error as the $L_2$ norm of the error in predicted stress history normalized by the $L_2$ norm of the stress history.  

The results are shown in Figure \ref{fig:2dfftresult}. Figure \ref{fig:2dfftresult}(a)  shows that the  test error (averaged over all test specimens) decreases with increasing amounts of training data and training epochs reaching an average of 5\% for a training size of 3200 in 400 epochs.  Figures \ref{fig:2dfftresult}(b,c) shows the input and output (both truth and approximation) for typical test and training samples with a neural net trained over 3200 samples and 500 epochs.   We conclude that our model reduction approach is able to learn a very accurate approximation of the map $\Psi$.

Figure \ref{fig:2dfftresult}(d) shows that we learn the behavior of the continuum model and is not tuned to any specific discretization used at the micro level.  For a given continuous input $U(t)$, we generate additional test samples (both input and output) with various time discretization from 50 to 1000 steps, and compare the error of the approximation trained with the original training set.  The figure shows that the error is independent of this resolution.    The figure also shows that the error depends on the dimension of the PCA reduction, decreasing with increasing dimension till it saturates at the dimension of the training data.  Finally, the figure shows the error due to PCA alone, and shows that the error of the learnt model is a few times that of the error of PCA.

Figures \ref{fig:2dfftresult}(e-g) show that the network trained using our protocol with random time steps provides a very accurate approximation of the map $\Psi$ in strain paths commonly encountered in practice.  These include uniaxial strain $U_{11} = f(t), U_{22}=1/f(t)$ and $U_{23}=0$ in Figure \ref{fig:2dfftresult}(e) for loading only when $f(t) = c t$,  $c \in (0, 1)$ and Figure \ref{fig:2dfftresult}(f) for loading unloading where $f(t) = c t, t \in (0,0.5); f(t) = c(t-0.5), t \in (0.5, 1)$ and for Figure \ref{fig:2dfftresult}(g) shear $U_{11} = (U_{22})^{-1} = 1/\sqrt{1+ct^2}$ (that we encounter in cavity expansion).

We use the same sampling of strain paths to obtain data in 3D, but using Taylor averaging.  We have conducted various tests of both the 2DFFT and 3DTaylor, and the results are gathered in \ref{app:tr}.  They lead us to conclude that using
random time steps is  effective even when tested against other strain paths, while using fixed time steps
only performs well on data generated with the same fixed time step but poorly on different data. Furthermore we conclude that the approximation error is independent of the rate exponent.

\paragraph{Causality and isotropy}

\begin{figure}
    \centering
    \includegraphics[width=4in]{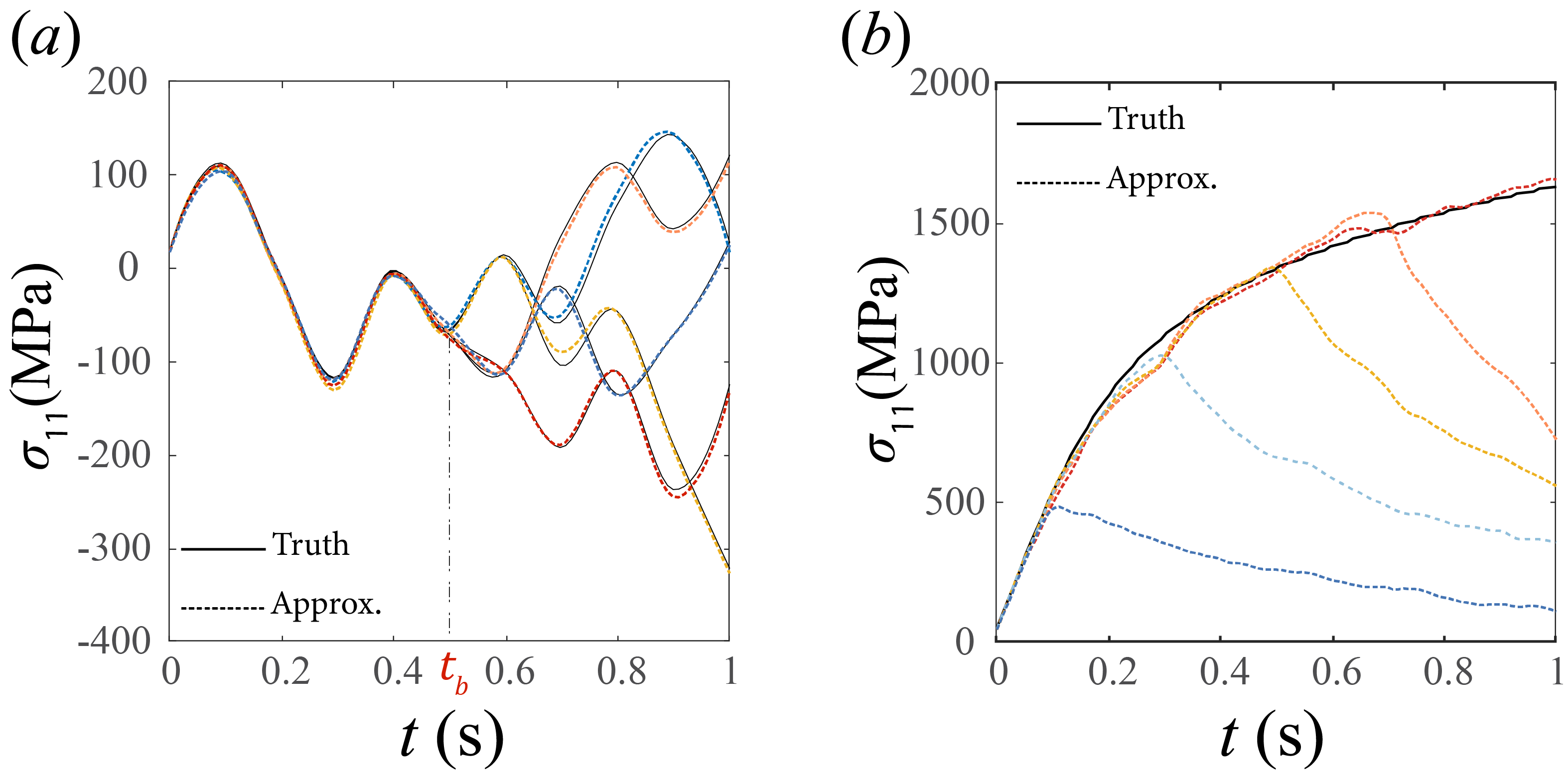}
    \caption{Causality.  (a) Test against five strain paths that diverge at $t=0.5$.  (b) Test against five strain paths that diverge at different instances of time. }
    \label{fig:causal}
\end{figure}

Recall that causality is essential to specify the necessary constitutive information (\ref{eq:fbar}) from mapping (\ref{eq:map}) that we approximate.
While our model and the resulting data is causal, the architecture we use is not restricted to be causal.  Figure  \ref{fig:causal} demonstrates that the approximation trained  from data that is causal automatically learns causality in the stress-strain
relationship.  We consider two sets of test samples.  In the first set, we consider five strain paths that are identical for $t \in (0,0.5)$ and distinct for $t \in (0.5,1)$.   We observe in Figure \ref{fig:causal}(a) that the approximation returns an identical stress response for $t \in (0,0.5)$ and distinct stress responses for $t \in (0.5,1)$.  In the second set, we consider five (uniaxial) strain paths where the strain increases linearly for  $t \in (0,t_i)$ for varying $t_i, i = 1, \dots 5$ and is then held constant.  Figure \ref{fig:causal}(b) shows that in all cases, the stress in the $i^{th}$ path follows identical paths to
those seeing the same data until time $t_i$, and then diverges.

We have chosen 128 randomly oriented grains to generate our 3DTaylor data, and therefore we expect the overall behavior to be almost isotropic.  Again our architecture does not impose this. However, the approximation automatically learns the isotropy from the data (see \ref{app:iso}).

\section{Application to impact problems}

We now use the trained neural network for macroscopic calculations of two classical impact problems, both in three dimensions.   We implement the net trained using the 3DTaylor unit cell calculations  (case 19 of Table \ref{tab:results}) as a material model (``VUMAT'') in the commercial finite element package ABAQUS \cite{manual2014abaqus}.  We emphasize that the neural network is only trained once for all the calculations presented below.

\subsection{Taylor anvil test}

\begin{figure}[t!]
    \centering
    \includegraphics[width=6in]{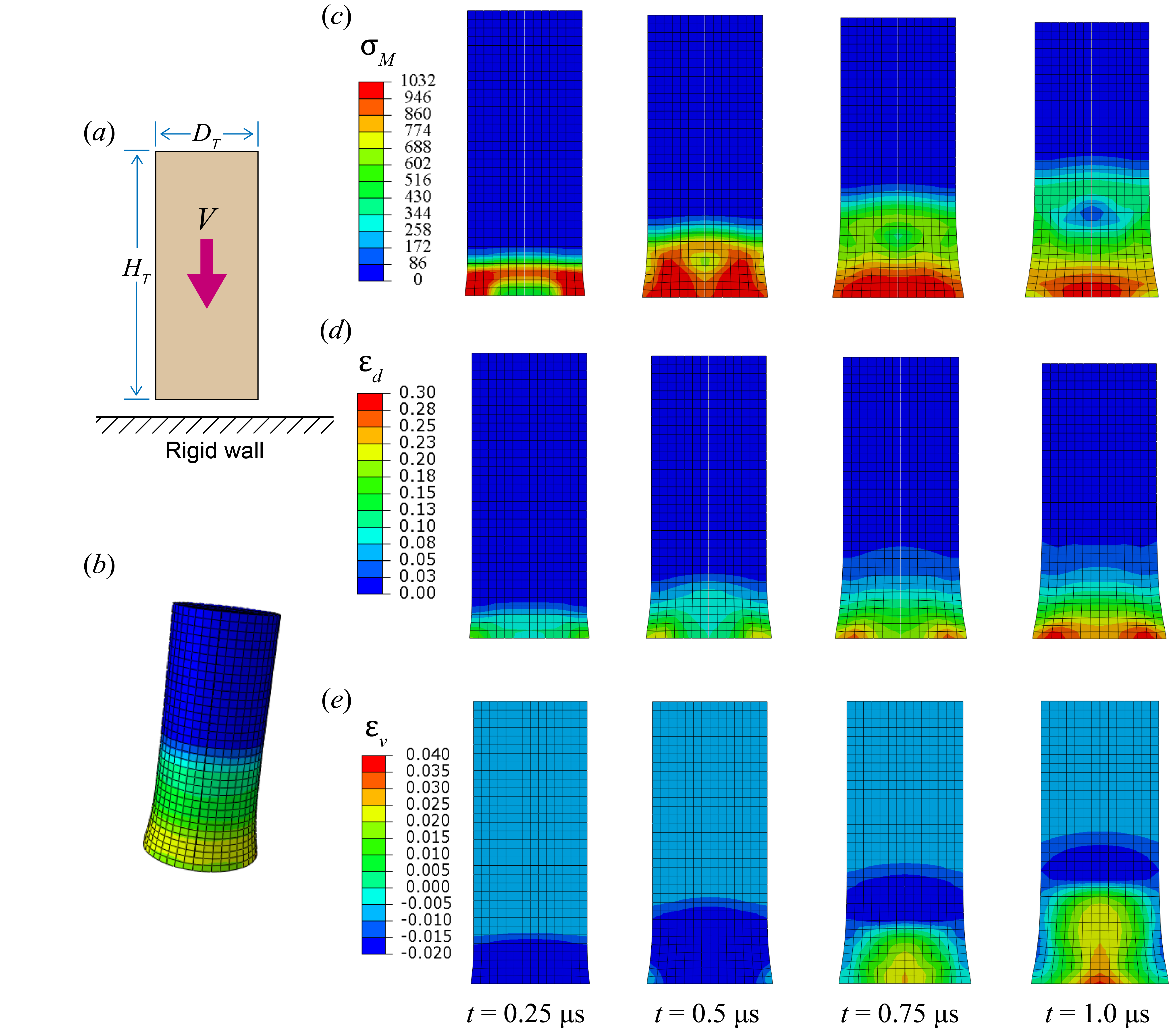}
    \caption{Typical results of a Taylor anvil test (a) Schematic.  (b) Deformation of the impactor at t = 1 $\mu$s.  (c-e) Snapshots an axial cross-section 	with the von Mises stress measure (c),  deviatoric strain measure (d) and volumetric strain measure (e).}
    \label{fig:taylor_anvil}
\end{figure}

A magnesium cylindrical impactor (of height $H_T$ = 5 mm and diameter $D_T$ = 1mm) traveling with an initial velocity $V$ = 200 m/s impacts a rigid friction-less wall at time $t = 0$ as shown schematically in Figure \ref{fig:taylor_anvil} (a).    
Figure \ref{fig:taylor_anvil}(c) shows the von Mises stress $\sigma_M = \sqrt{3/2}|\sigma -  (\text{tr} \sigma)/3  I|$ (where $|\cdot |$ denotes the Fr\"obenius matrix norm),  Figure \ref{fig:taylor_anvil}(d) the deviatoric strain measure $e_d = |F^T F /(\text{det} F)^2 - I |/2$ that indicates the evolution of the plastic deformation, and Figure \ref{fig:taylor_anvil}(e) the volumetric strain measure $e_v = \text{det}(F) - 1$ that indicates the longitudinal elastic wave.
Since the material is isotropic and geometry axisymmetric, the results are shown for an axial section.  Upon impact, an elastic wave first propagates into the impactor followed by a region of plastic deformation.  We also have a release wave moving in radially from the sides, leading to a complex radial distribution of the plastic deformation.

\begin{figure}[t]
    \centering
    \includegraphics[width=5.5in]{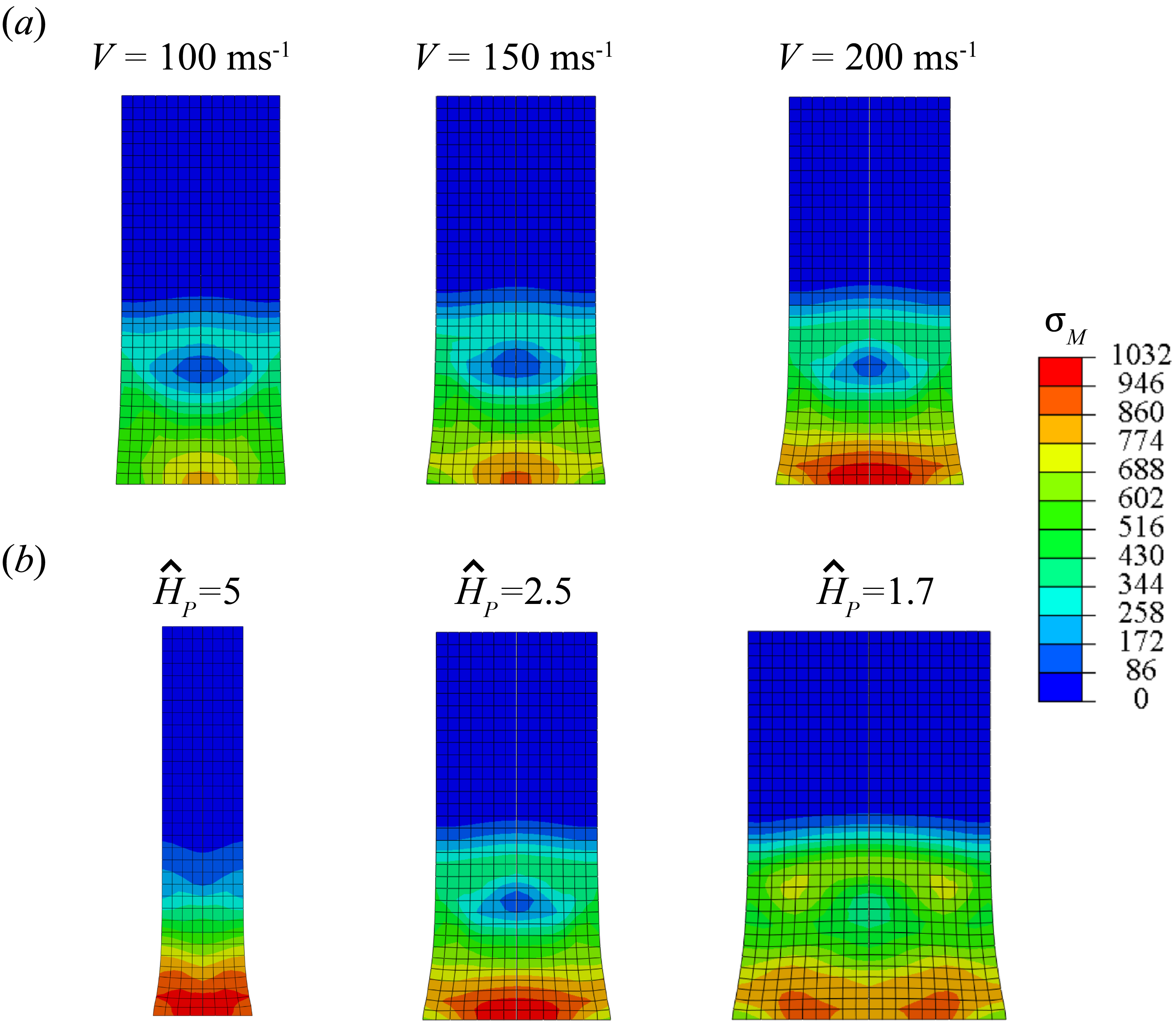}
    \caption{{\color{blue}} Parametric study of the Taylor anvil test.  Axial cross-section with the von Mises stress measure at $t_s=1$ for (a) different impact velocities $V$ and (b) various aspect ratios $\hat{H}_p = H_T/D_T$. }
    \label{fig:anvil_velocity}
\end{figure}

We repeat the calculation with various impact velocities and impactor geometries.   Figure \ref{fig:anvil_velocity} (a) shows the effect of the initial velocity $V$ on the deformation of the impactor. The elastic wave  propagates with similar velocity but increasing intensity, while the plastic deformation increases with increasing velocity.   The effect of changing the radius while keeping the height the same is shown in Fig. \ref{fig:anvil_velocity} (b)  for an impact velocity of $V$ = 200 m/s.  We see that the difference in the release wave from the sides changes the  stress and plastic deformation distributions.

\subsection{Projectile impact on a plate}

\begin{figure}[t]
    \centering
    \includegraphics[width=6.5in]{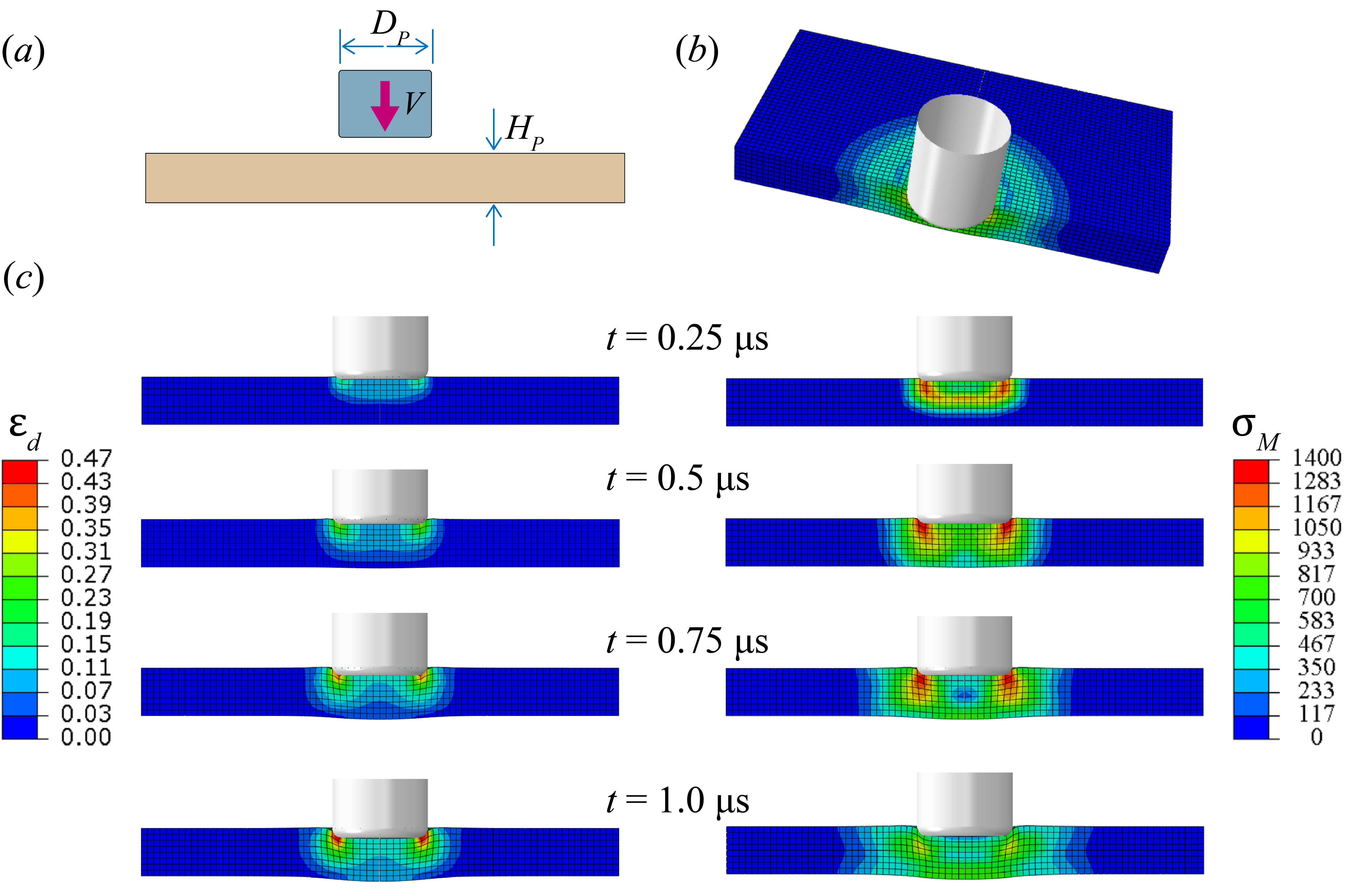}
    \caption{Typical results of a projectile impact on a plate (a) Schematic.  (b) Deformation in 3D.  (c-e) Snapshots an axial cross-section with the von Mises stress measure (c),  deviatoric strain measure (d) and volumetric strain measure (e). }
    \label{fig:plate}
\end{figure}

A cylindrical projectile of radius $D_p$ = 2 mm traveling at a velocity $V = $ 200 m/s impacts a large magnesium plate of thickness $H_p$ = 1 mm which is simply supported far away from the point of impact -- see Figure \ref{fig:plate}(a). The impacting cylinder is assumed to be rigid and infinitely dense compared to the plate and thus unaffected by the impact.  The deviatoric strain measure $e_d$ and the von Mises strain $\sigma_M$ are shown in Figure \ref{fig:plate}(c) and (d) respectively.  An elastic wave followed by a plastic wave propagates into the plate and is reflected from the free-face.  Subsequently, the plate becomes a wave guide with a radially expanding elastic and plastic waves as the impactor penetrates into the plate.

\begin{figure}[t]
    \centering
    \includegraphics[width=6.5in]{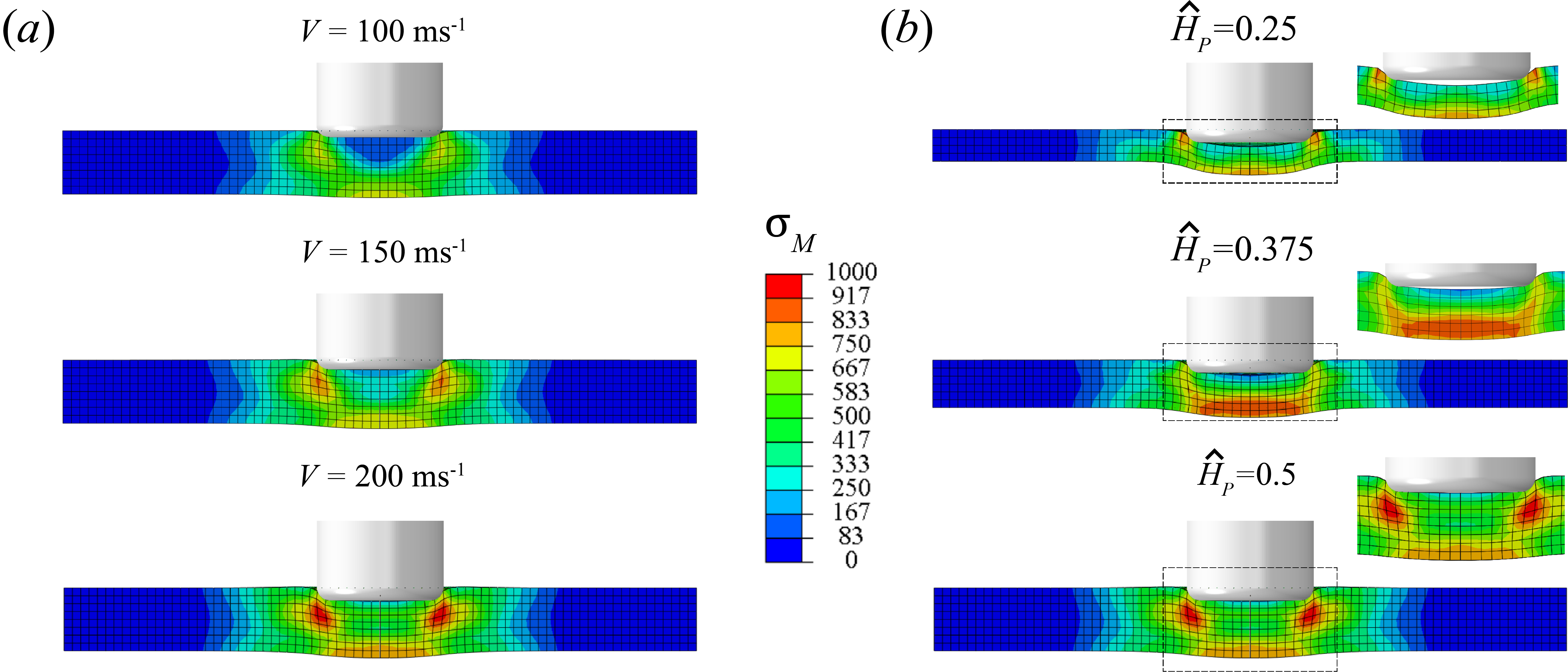}
    \caption{ Parametric study of the plate impact test.  Axial cross-section with von Mises stress measure at $t=1s$ for various (a) impact velocities $V$ and (b) plate thicknesses $\hat{H}_p=H_p/D_p$. }
    \label{fig:impact_thickness}
\end{figure}

Figure \ref{fig:impact_thickness} shows the results when we repeat the simulation with various impact velocities (Figure \ref{fig:impact_thickness} (a)) and plate thicknesses (Figure \ref{fig:impact_thickness} (b)).   The radially propagating elastic wave has the same velocity but increases in intensity with increasing impact velocity and decreasing plate thickness.  As the plate becomes thinner, we observe a change in the deformation mode with thin plates deforming in bending.  In particular, the plate begins to separate from the impactor at the center and the amount of plastic deformation is reduced.

\subsection{Computational effort}

\begin{table} \label{tab:cost}
\centering
\caption{Comparison of computation cost (wall-clock time in seconds)}
\includegraphics[width=6.5in]{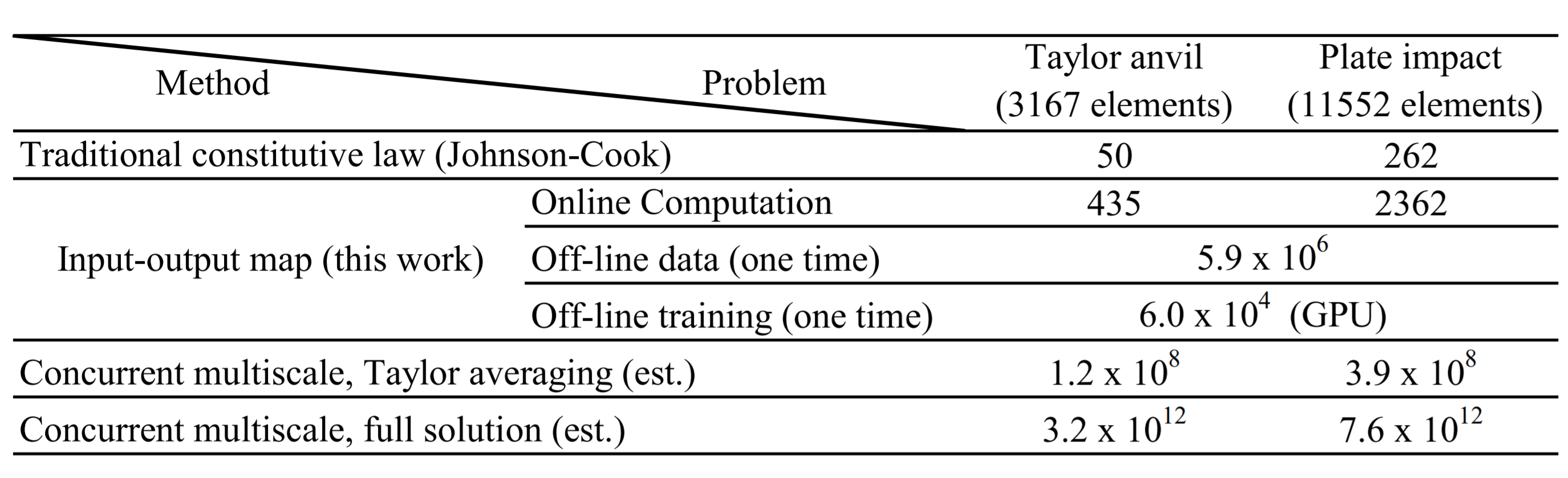}
\end{table}

The computational cost of the proposed method is compared with other approaches for the examples studied above in Table \ref{tab:cost}.  All calculations were preformed on a single core of Intel Skylake CPU (2.1 GHz) except the neural network training which was done on a NVIDIA P100 GPU with 3584 CUDA cores with a 1.3MHz clock (roughly, the Skylake CPU cost is bounded from above by 10$^3$ times the GPU cost).   We do not conduct concurrent multiscale simulations, but we may estimate the cost as a product of the cost of a unit cell problem, the number of elements and the number of time steps of the macroscopic problem.  

We find that the online cost of our method is only about ten times that of the computational cost of using an empirical constitutive law but orders of magnitude smaller than that of a concurrent multiscale method (with Taylor averaging).  Further, the one-time off-line cost of generating and training our approximation is comparable to that of a single calculation using a concurrent multiscale method.

\section{Discussion}

We have presented a framework that enables the specification of the macroscopic constitutive behavior (or closure relation) using microscopic computations combined with machine learning.   In this framework, data is generated by the solution of a fine scale model and is then used to train a surrogate which emulates the model's solution operator. The surrogate is then directly incorporated into coarse scale calculations.  We have demonstrated in problems involving the impact of polycrystalline inelastic solids that this approach can solve macroscopic problems with the fidelity of concurrent multiscale modeling and beyond (because we do not need a priori identification of state variables) at a few times the computational cost of solving the problem with an empirical model.

We demonstrated the framework using crystal plasticity, but it can easily be  extended to other continuum multiscale phenomena including composite materials, phase transitions, stress-assisted diffusion, and  discrete dislocation dynamics.  We can use our framework in an iterated manner to study hierarchical phenomena involving multiple scales \cite{liu2021hierarchical}.

We conclude with a discussion of ideas to build on this framework.  Our approach requires us to generate data over the entire interval of time to capture the memory of multiscale systems, and this may prove limiting in very long-time computations. However, in typical phenomena, this memory fades and therefore one may only need a limited history.  Such fading memory may be incorporated using a recurrent neural network (RNN) structure \cite{medsker2001recurrent}.  This eliminates the need to track the entire history, but potentially adds to the cost of training.   

A closely related question is whether RNN and related methods can use the data to discover an underlying Markovian macroscopic model by identifying appropriate macroscopic internal variables, delay kernels etc., which accurately describe the history-dependence. This would be exciting because it not only reduces computational cost but  provides new insights into the physics.

We have studied examples where the microscopic model can be studied in the long-wavelength limit so that the unit cell problem is over-damped.  The extension to situations like granular materials and molecular dynamics where inertia is critical at the microscale remains open.
%
%
%

Finally, our proposed approach requires training a new approximation for every starting material.  In our example of crystal plasticity, we need to generate data and perform training for every initial texture, every set of parameters in the single crystal model.  It will be useful if we can learn this as a part of the training, i.e., extend the map $\bar{\mathcal F}: \{ \{U(\tau): \tau \in (0,t)\}, \xi_0\} \} \mapsto \langle \sigma \rangle (t) \quad t \in (0,T)$.  This has been successfully demonstrated in simple problems like Darcy flow \cite{bhattacharya2020model}, and remains a work in progress.

\paragraph{Data availability}
The data and scripts needed to evaluate the conclusions of this paper are available in the GitHub repository ``Learning based multiscale
" (\url{https://github.com/Burigede/Learning_based_multiscale.git}). 

\paragraph{Author contributions}
BL performed all the numerical simulations.   BL, NK, AS and KB took the lead in identifying the problem and the initial formulation.  All authors were involved in the detailed formulation and in analyzing the results.  BL and KB made the initial draft which all authors revised.

\paragraph{Author interests}
The authors declare no competing interest.

\paragraph{Acknowledgement}

We are grateful to Dennis Kochmann for discussion and for providing us with the 2DFFT and the 3D Taylor code to generate the data.  This research was sponsored by the Army Research Laboratory and was accomplished under Cooperative Agreement Number W911NF-12-2-0022. The views and conclusions contained in this document are those of the authors and should not be interpreted as representing the official policies, either expressed or implied, of the Army Research Laboratory or the U.S. Government. The U.S. Government is authorized to reproduce and distribute reprints for Government purposes notwithstanding any copyright notation herein.
ZL is supported by the Kortschak Scholars Program.
AS is supported in part by Bren endowed chair and De Logi grant. 
AMS is also partially supported by NSF grant DMS-1818977

\bibliographystyle{abbrv}
\bibliography{learnt_multiscale.bib}

\begin{thebibliography}{10}

\bibitem{manual2014abaqus}
{ABAQUS} user manual: {ABAQUS} theory guide, 2014.

\bibitem{asaro1983crystal}
R.~J. Asaro.
\newblock {Crystal plasticity}.
\newblock {\em Journal of Applied Mechanics}, 50:921--934, 12 1983.

\bibitem{balasubramanian2002elasto}
S.~Balasubramanian and L.~Anand.
\newblock Elasto-viscoplastic constitutive equations for polycrystalline fcc
  materials at low homologous temperatures.
\newblock {\em Journal of the Mechanics and Physics of Solids}, 50:101--126,
  2002.

\bibitem{betal_jap11}
N.~R. Barton, J.~V. Bernier, R.~Becker, A.~Arsenlis, R.~Cavallo, J.~Marian,
  M.~Rhee, H.-S. Park, B.~A. Remington, and R.~T. Olson.
\newblock A multiscale strength model for extreme loading conditions.
\newblock {\em Journal of Applied Physics}, 109:073501, 2011.

\bibitem{bensoussan2011asymptotic}
A.~Bensoussan, J.-L. Lions, and G.~Papanicolaou.
\newblock {\em Asymptotic analysis for periodic structures}, volume 374.
\newblock American Mathematical Soc., 2011.

\bibitem{b_prs_99}
K.~Bhattacharya.
\newblock {Phase boundary propagation in a heterogeneous body}.
\newblock {\em Proceedings of the Royal Society of London A}, 455:757--766,
  1999.

\bibitem{bhattacharya2020model}
K.~Bhattacharya, B.~Hosseini, N.~B. Kovachki, and A.~M. Stuart.
\newblock Model reduction and neural networks for parametric pdes.
\newblock {\em ArXiv e-prints}, 2005.03180, 2020.

\bibitem{bc_book_13}
V.~Bulatov and W.~Cai.
\newblock {\em Computer Simulations of Dislocations}.
\newblock Oxford University Press, 2013.

\bibitem{car1985unified}
R.~Car and M.~Parrinello.
\newblock Unified approach for molecular dynamics and density-functional
  theory.
\newblock {\em Physical Review Letters}, 55:2471, 1985.

\bibitem{Chang2015AMagnesium}
Y.~Chang and D.~M. Kochmann.
\newblock {A variational constitutive model for slip-twinning interactions in
  hcp metals: Application to single- and polycrystalline magnesium}.
\newblock {\em International Journal of Plasticity}, 73:39--61, 8 2015.

\bibitem{cd_mse_18}
Z.~Chen and S.~Daly.
\newblock Deformation twin identification in magnesium through clustering and
  computer vision.
\newblock {\em Materials Science and Engineering: A}, 736:61--75, 2018.

\bibitem{cheng2019first}
T.~Cheng, A.~Jaramillo-Botero, Q.~An, D.~V. Ilyin, S.~Naserifar, and W.~A.
  Goddard.
\newblock First principles-based multiscale atomistic methods for input into
  first principles nonequilibrium transport across interfaces.
\newblock {\em Proceedings of the National Academy of Sciences},
  116:18193--18201, 2019.

\bibitem{cole2020machine}
D.~J. Cole, L.~Mones, and G.~Cs{\'a}nyi.
\newblock A machine learning based intramolecular potential for a flexible
  organic molecule.
\newblock {\em Faraday Discussions}, 224:247--264, 2020.

\bibitem{collobert2008unified}
R.~Collobert and J.~Weston.
\newblock A unified architecture for natural language processing: Deep neural
  networks with multitask learning.
\newblock In {\em Proceedings of the 25th international conference on Machine
  learning}, pages 160--167, 2008.

\bibitem{dr_book_11}
R.~de~Borst and E.~Ramm.
\newblock {\em Multiscale Methods in Computational Mechanics}.
\newblock Springer, Heidelberg, 2011.

\bibitem{m_mse_20}
M.~De~Graef.
\newblock A dictionary indexing approach for {EBSD}.
\newblock {\em Materials Science and Engineering}, 891:012009, 2020.

\bibitem{weinan2011principles}
W.~E.
\newblock {\em Principles of multiscale modeling}.
\newblock Cambridge University Press, 2011.

\bibitem{feyel2000fe2}
F.~Feyel and J.-L. Chaboche.
\newblock Fe2 multiscale approach for modelling the elastoviscoplastic
  behaviour of long fibre sic/ti composite materials.
\newblock {\em Computer methods in applied mechanics and engineering},
  183:309--330, 2000.

\bibitem{f_book_10}
M.~Finnis.
\newblock {\em Crystals, defects and microstructures: Modeling across scales}.
\newblock Oxford University Press, 2010.

\bibitem{f_book_09}
E.~Fish.
\newblock {\em Multiscale Methods: Bridging the Scales in Science and
  Engineering}.
\newblock Oxford University Press, Oxford, 2009.

\bibitem{fritzen2009periodic}
F.~Fritzen, T.~B{\"o}hlke, and E.~Schnack.
\newblock Periodic three-dimensional mesh generation for crystalline aggregates
  based on voronoi tessellations.
\newblock {\em Computational Mechanics}, 43:701--713, 2009.

\bibitem{fu2005multiscale}
C.-C. Fu, J.~Dalla~Torre, F.~Willaime, J.-L. Bocquet, and A.~Barbu.
\newblock Multiscale modelling of defect kinetics in irradiated iron.
\newblock {\em Nature materials}, 4:68--74, 2005.

\bibitem{g_book_14}
F.~Giustino.
\newblock {\em Materials Modelling using Density Functional Theory: Properties
  and Predictions}.
\newblock Oxford University Press, 2014.

\bibitem{goldberg2017neural}
Y.~Goldberg.
\newblock Neural network methods for natural language processing.
\newblock {\em Synthesis lectures on human language technologies}, 10:1--309,
  2017.

\bibitem{gfa_book_13}
M.~Gurtin, E.~Fried, and L.~Anand.
\newblock {\em The Mechanics and Thermodynamics of Continua}.
\newblock Oxford University Press, 2013.

\bibitem{he2016deep}
K.~He, X.~Zhang, S.~Ren, and J.~Sun.
\newblock Deep residual learning for image recognition.
\newblock In {\em Proceedings of the IEEE conference on computer vision and
  pattern recognition}, pages 770--778, 2016.

\bibitem{jetal_aplm_13}
A.~Jain, S.~P. Ong, G.~Hautier, W.~Chen, W.~D. Richards, S.~Dacek, S.~Cholia,
  D.~Gunter, D.~Skinner, and G.~Ceder.
\newblock { The Materials Project: A materials genome approach to accelerating
  materials innovation}.
\newblock {\em APL Materials}, 1:011002, 2013.

\bibitem{jordan2020neural}
B.~Jordan, M.~B. Gorji, and D.~Mohr.
\newblock Neural network model describing the temperature-and rate-dependent
  stress-strain response of polypropylene.
\newblock {\em International Journal of Plasticity}, 135:102811, 2020.

\bibitem{kd_armr_15}
S.~Kalidindi and M.~De~Graef.
\newblock Materials data science: current status and future outlook.
\newblock {\em Annual Review of Materials Research}, 45:171--193, 2015.

\bibitem{Klambauer2017Self-normalizingNetworks}
G.~Klambauer, T.~Unterthiner, A.~Mayr, and S.~Hochreiter.
\newblock {Self-normalizing neural networks}.
\newblock In {\em Advances in neural information processing systems}, pages
  971--980, 2017.

\bibitem{ketal_book_00}
U.~Kocks, C.~Tome, and H.-R. Wenk.
\newblock {\em Texture and Anisotropy}.
\newblock Cambridge University Press, 2000.

\bibitem{lecun1995convolutional}
Y.~LeCun, Y.~Bengio, et~al.
\newblock Convolutional networks for images, speech, and time series.
\newblock {\em The handbook of brain theory and neural networks},
  3361(10):1995, 1995.

\bibitem{liu2021hierarchical}
B.~Liu, X.~Sun, K.~Bhattacharya, and M.~Ortiz.
\newblock Hierarchical multiscale quantification of material uncertainty.
\newblock {\em arXiv preprint arXiv:2102.02927}, 2021.

\bibitem{liu2019deep1}
Z.~Liu, C.~Wu, and M.~Koishi.
\newblock A deep material network for multiscale topology learning and
  accelerated nonlinear modeling of heterogeneous materials.
\newblock {\em Computer Methods in Applied Mechanics and Engineering},
  345:1138--1168, 2019.

\bibitem{l_ncm_19}
A.~Ludwig.
\newblock Discovery of new materials using combinatorial synthesis and
  high-throughput characterization of thin-film materials libraries combined
  with computational methods.
\newblock {\em npj Computational Materials}, 5:70, 2019.

\bibitem{marchand2020machine}
D.~Marchand, A.~Jain, A.~Glensk, and W.~Curtin.
\newblock Machine learning for metallurgy i. a neural-network potential for
  al-cu.
\newblock {\em Physical Review Materials}, 4(10):103601, 2020.

\bibitem{medsker2001recurrent}
L.~R. Medsker and L.~Jain.
\newblock Recurrent neural networks.
\newblock {\em Design and Applications}, 5, 2001.

\bibitem{Mozaffar2019DeepPlasticity}
M.~Mozaffar, R.~Bostanabad, W.~Chen, K.~Ehmann, J.~Cao, and M.~A. Bessa.
\newblock {Deep learning predicts path-dependent plasticity}.
\newblock {\em Proceedings of the National Academy of Sciences of the United
  States of America}, 116:26414--26420, 2019.

\bibitem{pavliotis2008multiscale}
G.~Pavliotis and A.~Stuart.
\newblock {\em Multiscale methods: averaging and homogenization}.
\newblock Springer Science \& Business Media, 2008.

\bibitem{p_book_01}
R.~Phillips.
\newblock {\em Crystals, defects and microstructures: Modeling across scales}.
\newblock Cambridge University Press, 2001.

\bibitem{rice1971inelastic}
J.~R. Rice.
\newblock Inelastic constitutive relations for solids: an internal-variable
  theory and its application to metal plasticity.
\newblock {\em Journal of the Mechanics and Physics of Solids}, 19:433--455,
  1971.

\bibitem{tadmor1996quasicontinuum}
E.~B. Tadmor, M.~Ortiz, and R.~Phillips.
\newblock Quasicontinuum analysis of defects in solids.
\newblock {\em Philosophical magazine A}, 73:1529--1563, 1996.

\bibitem{umehara2019analyzing}
M.~Umehara, H.~S. Stein, D.~Guevarra, P.~F. Newhouse, D.~A. Boyd, and J.~M.
  Gregoire.
\newblock Analyzing machine learning models to accelerate generation of
  fundamental materials insights.
\newblock {\em npj Computational Materials}, 5:34, 2019.

\bibitem{van2020roadmap}
E.~Van Der~Giessen, P.~A. Schultz, N.~Bertin, V.~V. Bulatov, W.~Cai,
  G.~Cs{\'a}nyi, S.~M. Foiles, M.~G. Geers, C.~Gonz{\'a}lez, M.~H{\"u}tter,
  et~al.
\newblock Roadmap on multiscale materials modeling.
\newblock {\em Modelling and Simulation in Materials Science and Engineering},
  28:043001, 2020.

\bibitem{vidyasagar2017predicting}
A.~Vidyasagar, W.~L. Tan, and D.~M. Kochmann.
\newblock Predicting the effective response of bulk polycrystalline
  ferroelectric ceramics via improved spectral phase field methods.
\newblock {\em Journal of the Mechanics and Physics of Solids}, 106:133--151,
  2017.

\bibitem{wen2019hybrid}
M.~Wen and E.~B. Tadmor.
\newblock Hybrid neural network potential for multilayer graphene.
\newblock {\em Physical Review B}, 100:195419, 2019.

\bibitem{wold1987principal}
S.~Wold, K.~Esbensen, and P.~Geladi.
\newblock Principal component analysis.
\newblock {\em Chemometrics and intelligent laboratory systems}, 2:37--52,
  1987.

\bibitem{xiao2019machine}
S.~Xiao, R.~Hu, Z.~Li, S.~Attarian, K.-M. Bj{\"o}rk, and A.~Lendasse.
\newblock A machine-learning-enhanced hierarchical multiscale method for
  bridging from molecular dynamics to continua.
\newblock {\em Neural Computing and Applications}, pages 1--15, 2019.

\end{thebibliography}

\newpage
\begin{center}
A learning-based multiscale method and its application to inelastic impact problems\\
Liu,  Kovachki, Li, Azizzadenesheli, Anandkumar, Stuart, Bhattacharya\\
\vspace{\baselineskip}
{\Huge \bf Supplementary Information}
\vspace{\baselineskip}
\end{center}
\setcounter{section}{0}
\setcounter{page}{1}
\setcounter{table}{0}
\setcounter{figure}{0}

\renewcommand{\thesection}{SI-\Alph{section}}
\renewcommand{\thepage}{SI-\arabic{page}}
\renewcommand{\thetable}{SI-\arabic{table}}
\renewcommand{\thefigure}{SI-\arabic{figure}}

\section{Crystal plasticity with twinning} \label{app:cp}

We consider the constitutive framework developed by Chang and Kochmann \cite{Chang2015AMagnesium}.  The internal variables are the crystallographic orientation $Q \in SO(d)$, total inelastic deformation gradient $F_\text{in} \in {\mathbb R}^{d\times d}$, the slip activity $\gamma = \{\gamma_\alpha\}_{\alpha=1}^{n_s}$ in the $n_s$ slip systems and twin volume fractions $\lambda = \{\lambda_\beta \}_{\beta=1}^{n_t}$ in the $n_t$ twin systems that satisfy $\lambda_\beta \in [0,1], \ \sum_\beta \lambda_\beta = 1$.
We introduce secondary internal variables (accumulated plastic activity) $\{ e_\alpha \}_{\alpha=1}^{n_s}$ by integrating
\begin{equation}
\dot{e}_\alpha = | \dot{\gamma}_\alpha|.
\end{equation}

To specify the constitutive functions $S$ and $K$, we specify a stored energy density 
\begin{align}
&W(F,F_\text{in},e,\lambda,y) = W_e(F F_\text{in}^{-1}) + W_p(e) + W_t(\lambda) \quad \text{where}\\
&\quad \quad W_e(A)  = {G \over 2} \left( {\text{tr} A^T A \over (\det A)^{2/3}} - 3 \right) + \lambda_e (\det A - 1)^2,\\
&\quad \quad W_p(e) = \frac{1}{2}e \cdot \mathcal{H} e+\sum_{\alpha = 1}^{n_s}\sigma_{\alpha}^{\infty}(e_{\alpha} + \frac{\sigma_{\alpha}^{\infty}}{h_{\alpha}} \text{exp}(\frac{-h_{\alpha}e_{\alpha}}{\sigma_{\alpha}^{\infty}})), \\
&\quad \quad W_t(\lambda) = \lambda \cdot \mathcal{K} \lambda + \sum_{\beta = 1}^{n_t}\frac{1}{2}h_{\beta}\lambda_{\beta}^{2}
\end{align}
and dissipation functions
\begin{align}
&D_p( \dot{\gamma}) = \sum_{\alpha=1}^{n_s}\frac{\tau_{0,\alpha}\dot{\gamma}_{0,\alpha}}{m_\alpha+1}(\frac{\dot{\gamma}_\alpha}{\dot{\gamma}_{0,\alpha}})^{m_\alpha+1}, \quad \\
&D_t (\dot{\lambda}) = \sum_{\beta=1}^{n_t}\frac{\tau_{0,\beta}\dot{\lambda}_{0,\beta}}{m_\beta+1}(\frac{\dot{\lambda}_\beta}{\dot{\lambda}_{0,\beta}})^{m_\beta+1}.
\end{align}

The stress function is specified as
\begin{align}
S(F,F_\text{in},e,\lambda) = {\partial W \over \partial F} (F,F_\text{in},e,\lambda) = 
\left. {\partial W_e \over \partial A} \right|_{F,F_\text{in},e,\lambda,y)} F_\text{in}^{-1}.
\end{align}
Note that we have chosen an isotropic elastic law for convenience and it does not explicitly depend on position $y$.

The kinetic relations are specified as
\begin{align}
& \dot{F}_\text{in} F_\text{in}  = L_p + L_t \quad \text{where}\\
& \quad \quad L_p = \sum_{\alpha = 1}^{n_s}\dot{\gamma}_{\alpha}[(1-\sum_{\beta=1}^{n_t}\lambda_{\beta})s_{\alpha} \otimes m_{\alpha} + \sum_{\beta=1}^{n_t} \lambda_{\beta} s_{\alpha\beta} \otimes m_{\alpha\beta}],\\
& \quad \quad L_t = \gamma_{t}\sum_{\beta=1}^{n_t} \dot{\lambda}_{\beta} s_{\beta} \otimes m_{\beta},\\
& 0 \in {\partial \over \partial \gamma_\alpha} (W + D_p),\\
& 0 \in {\partial \over \partial \lambda_\beta} (W + D_t).
\end{align}
Crucially $s_\alpha, m_\alpha, b_\alpha, m_\alpha$ depend on position.
Note that the final two kinetic relations are written as differential inclusions because the derivative is not smooth ($W_p$ is specified in terms of $e$).

In the 2D calculations, we consider two orthogonal slip systems $s_1 = (1,0,0), m_1 = (0,1,0)$ and $s_2 = (0,1,0), m_2 = (0,-1,0)$ and no twinning. The shear modulus $G$ and Lame's constant $\lambda$ is chosen to be 19GPa and 24GPa respectively, while the initial yield strength $\tau_0$ and reference slip rate $\gamma_0$ is 100MPa and 1$\text{s}^{-1}$. The strain rate sensitivity are chosen to be $m = 0.5$. 

We list the parameters that we use for the 3D calculations in Table \ref{tab:Taylorparam}. 

\begin{table}[t]
\centering
\caption{Parameters used for magnesium in the crystal-plasticity model.}
\begin{tabular}{l l l l}
\hline
\hline
& Parameter & Value & Unit\\
\hline
\multirow{5}{*}{Basal $\langle a \rangle$}
& $h_{\alpha}$ & 7.1 &GPa\\
& $\sigma_{\alpha}^{\infty}$ & 0.7 &MPa\\
& $h_{ij}$ & 0 &MPa \\
& $m_{\alpha}$ & 0.5 & -\\
& $\dot{\gamma}_{0,\alpha}$ & 1.0$\cdot 10^5$ &$\text{s}^{-1}$\\
\hline
\multirow{5}{*}{Prismatic $\langle a \rangle$}
& $h_{\alpha}$ & 40 &GPa\\
& $\sigma_{\alpha}^{\infty}$ & 170 &MPa\\
& $h_{ij}$ & 20 &MPa \\
& $m_{\alpha}$ & 0.5 & -\\
& $\dot{\gamma}_{0,\alpha}$ & 1.0$\cdot 10^5$ &$\text{s}^{-1}$\\
\hline
\multirow{6}{*}{Pyramidal $\langle c+a \rangle$}
& $h_{\alpha}$ & 30 &GPa\\
& $\sigma_{\alpha}^{\infty}$ & 200 &MPa\\
& $h_{ij}$ & 25 &MPa \\
& $\tau_{0,\alpha}$ & 50.5 &MPa\\
& $m_{\alpha}$ & 0.5 & -\\
& $\dot{\gamma}_{0,\alpha}$ & 1.0$\cdot 10^5$ &$\text{s}^{-1}$\\
\hline
\multirow{5}{*}{Tensile twin}
& $h_{\beta}$ & 7 &MPa \\
& $k_{ij}$ & 40 &GPa \\
& $m_{\beta}$ & 0.5 & -\\
& $\dot{\lambda}_{0,\beta}$ & 1.0$\cdot 10^5$ &$\text{s}^{-1}$ \\
& $\gamma_t$ &  0.129 & -\\
\hline
\multirow{2}{*}{Elastic Lame Moduli}
& $\lambda_{e}$ & 24 &GPa \\
& $G$ & 25 &GPa \\
\hline
\multirow{1}{*}{Density}
& $\rho$ & $1.0\cdot 10^4$ &$\text{kg}$ $\text{m}^{-3}$  \\
\hline
\end{tabular}
\label{tab:Taylorparam}
\end{table}

\section{Various tests of training} \label{app:tr}

\begin{sidewaystable}
    \centering
    \caption{Table of all training we have conducted.}
    \includegraphics[width=0.9\columnwidth]{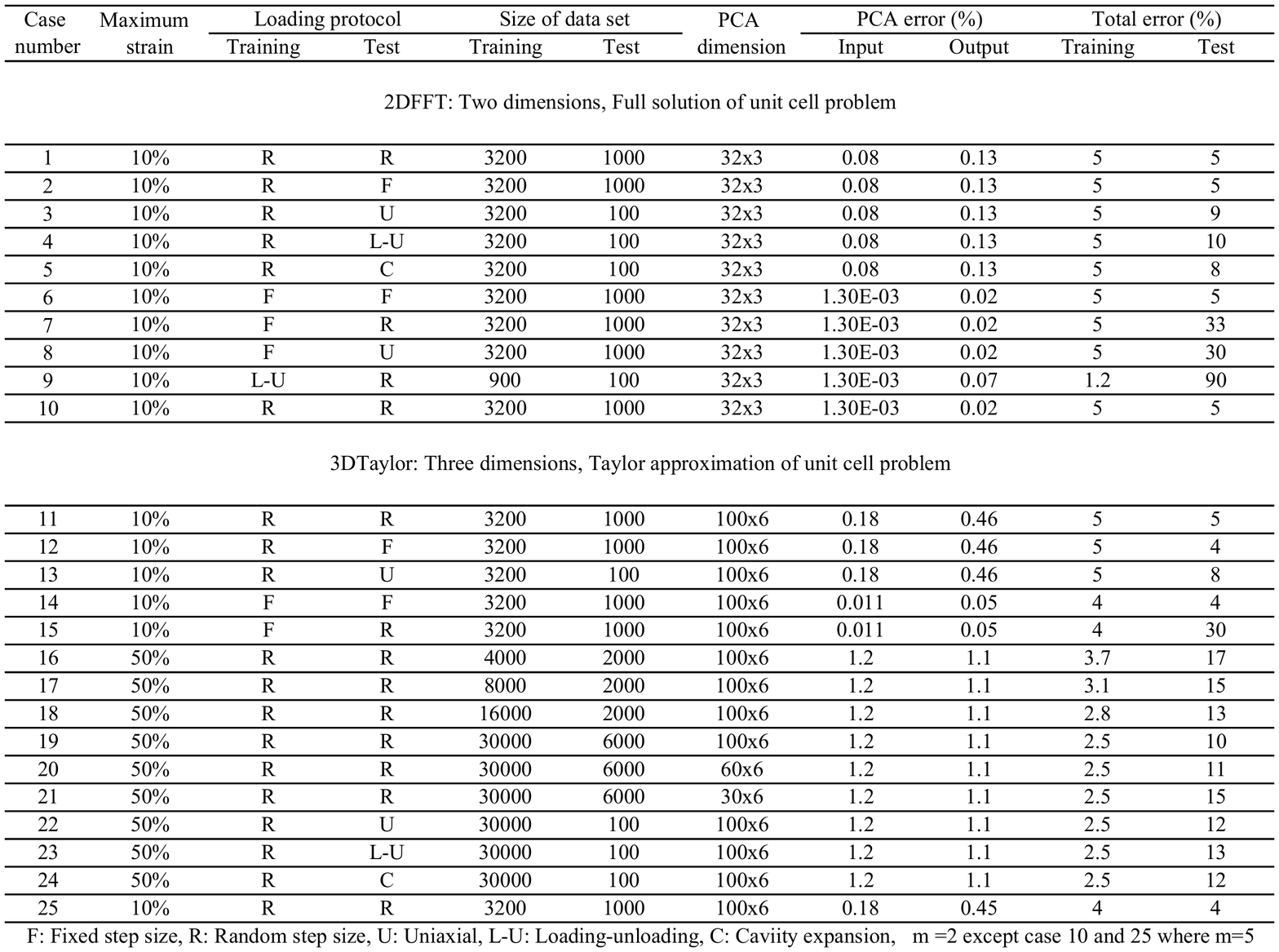}
    \label{tab:results}
\end{sidewaystable}

We have conducted various tests of both the 2DFFT and 3DTaylor and the results are gathered in Table \ref{tab:results}.  Cases 1-5, 11-12, 22-24 show that training with random time-steps is  effective even when tested against other strain paths in both 2DFFT and 3DTaylor.  Fixed time step does well against data with fixed time step but poorly against other data (case 5-7. 13-14).  Comparison between case 11 and 19 shows that we need more training data as the maximum strain increases.  Cases 19-21 shows that increasing the PCA dimension reduces the error.  Finally cases 9 and 23 show that the approximation error is independent of the rate exponent.

\section{Isotropy} \label{app:iso}

\begin{figure}
    \centering
    \includegraphics[width=4in]{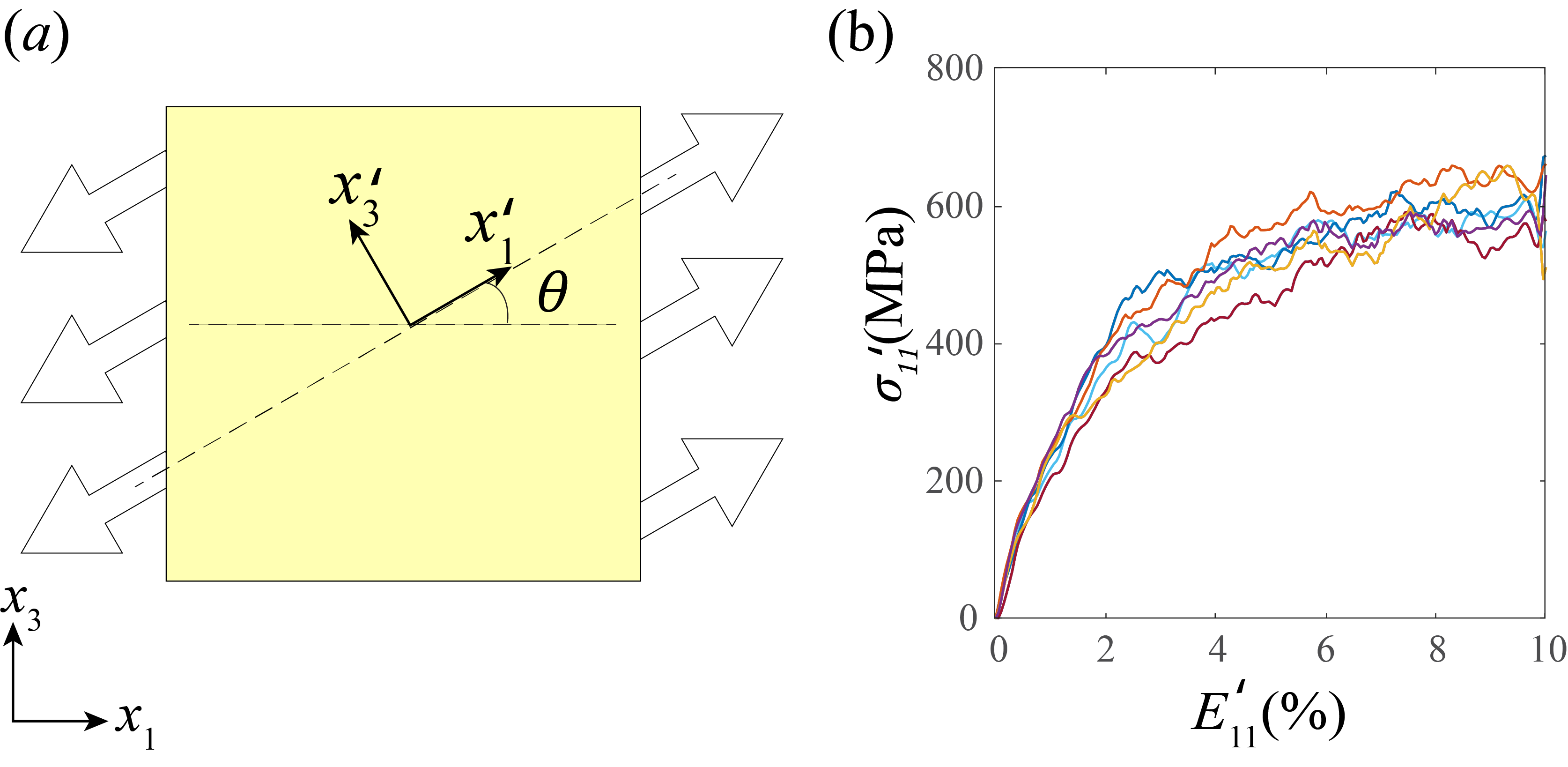}
    \caption{Isotropy.  (a) Schematic of the boundary condition.  (b) Tensile stress at local coordinate with $\theta$ in the range of $[0,30^{\circ},60^{\circ}, ..., 360^{\circ}]$.}
    \label{fig:isotropy}
\end{figure}

It is well known that a polycrystalline solid with a random initial grain orientation is isotropic on the macroscale. Here we demonstrate that the learned model is capable of capturing the isotropic macroscopic behavior through data. We test the neural network trained with the 3D Taylor model (case 19 in Table \ref{tab:results}) by applying uniaxial tensile loading in the local (material coordinate) $x_1'-x_3'$ formulated by transforming the global coordinate system $x_1-x_3$ with an angle $\theta$ perpendicular to the $x_2$ direction. The computed tensile stress-strain curve in the material coordinate $x_1'-x_3'$ is reported in Figure \ref{fig:isotropy}. The results show that the trained neural net can capture the macroscopic isotropy as the tensile stress remains approximately the same regardless of the direction of the tensile loading. It bears to note that we do not enforce the principle of isotropy in our neural network model but is learned through the training. 

\end{document}